\documentclass[12pt,preprint]{aastex}

\shorttitle{Connection between Superluminal Ejections and Gamma-Ray Flares}
\shortauthors{Jorstad et al.}

\begin{document}

\title{Multi-Epoch VLBA Observations of EGRET-Detected Quasars and
BL Lac Objects: Connection between Superluminal Ejections and Gamma-Ray 
Flares in Blazars} 
    
\author{Svetlana G. Jorstad\altaffilmark{1,2,3}, Alan P. Marscher\altaffilmark{1}, John R. Mattox\altaffilmark{1,4}}
\author{Margo F. Aller\altaffilmark{5}, Hugh D. Aller,\altaffilmark{5}} 
\author{Ann E. Wehrle\altaffilmark{6},\and Steven D. Bloom,\altaffilmark{7}}
\altaffiltext{1}{Institute for Astrophysical Research, Boston University, 
      725 Commonwealth Ave., Boston, MA, 02215}
\altaffiltext{2}{Astronomical Institute, St. Petersburg State University,
       Bibliotechnaya pl. 2, Petrodvorets, St. Petersburg, 198904,
       Russia}
\altaffiltext{3}{Formerly S. G. Marchenko}
\altaffiltext{4}{Current address: Department of Chemistry, Physics and Astronomy,
Francis Marion University, Florence, SC 29501-0547}
\altaffiltext{5}{Astronomy Department, University of Michigan, Ann Arbor, 
MI 48109}
\altaffiltext{6}{Jet Propulsion Laboratory, MS 301-486, 4800 Oak Grove Dr.,
Pasadena, CA 91109}
\altaffiltext{7}{Hampden-Sydney College, Box 821, Hampden-Sydney, VA 23943}

\slugcomment{\apj, 01/26/01}
\begin{abstract}
We examine the coincidence of times of high $\gamma$-ray flux
and ejections of superluminal components from the core
in EGRET blazars based on a VLBA monitoring program at 22 and 43 GHz
from November 1993 to July 1997.  In 23 cases of $\gamma$-ray flares for which sufficient
VLBA data exist, 10 of the flares (in 8 objects) fall within 1$\sigma$ uncertainties of
the extrapolated epoch of zero separation from the core of a superluminal
radio component. In each of two sources (0528$+$134 and 1730$-$130)  
two successive $\gamma$-ray flares were followed by the appearance of
new superluminal components. We carried out statistical simulations which
show that if the number of coincidences $\ge$ 10 the radio and $\gamma$-ray events
are associated with each other at $>$99.999\% confidence. 
Our analysis of the observed behavior, including variability
of the polarized radio flux, of the sources before, during, and after the
$\gamma$-ray flares suggests 
that the $\gamma$-ray events occur in the superluminal radio knots.
This implies that the $\gamma$-ray
flares are caused by inverse Compton scattering
by relativistic electrons in the parsec-scale regions of the jet rather than
closer to the central engine.
\end{abstract}


\section{Introduction}

We have completed a monitoring program of the milliarcsecond-scale
structure of $\gamma$-ray bright blazars (42 sources) with the VLBA at 22 and
43~GHz during the period November 1993 to July 1997. The images, model fits,
and basic statistical analysis of this study are presented in Jorstad et al. (2001).
Based on  proper
motions in 33 sources, we have found that the apparent superluminal motions in $\gamma$-ray 
sources are much faster than for the general population of bright compact radio sources.
The results strongly support
the thesis that the $\gamma$-ray emission originates in a highly
relativistic jet. A positive correlation (coefficient of correlation 0.45)
between VLBI core flux and
$\gamma$-ray flux suggests that the production of the $\gamma$-ray 
emission takes place in the most compact regions of the relativistic jet
(Mattox et al. 1997a; Jorstad et al. 2001).
Comparison of Mets\"ahovi radio (22 and 37~GHz) total flux density variations
and EGRET $\gamma$-ray observations
led L\"ahteenm\"aki, Valtaoja, \& Tornikoski (1999) to suggest that the radio and 
$\gamma$-ray emission 
originates within the same shocked area of the relativistic jet and that the $\gamma$-rays
are most likely produced by the synchrotron self-Compton mechanism. 
If this is the case, then  VLBI component ejections
should correspond to outbursts in the $\gamma$-ray light curves. 
The first case in which the ejection of a jet component 
was contemporaneous with enhanced level 
of $\gamma$-ray flux was reported in the quasar 3C~279 by Wehrle, Zoock, \& Unwin (1994). 
Several other cases were found during the EGRET mission: 
3C~454.3 (Krichbaum et al. 1995),  
3C~273 (Krichbaum et al. 1996), 0836$+$710
(Otterbein et al. 1998), 1611$+$343 (Piner \& Kingham 1998), and 0528$+$134 (Britzen et al. 1999). 
On the other hand,
no components associated  with the $\gamma$-ray flares of CTA~26,
1156$+$295, and 1606$+$106 (Piner \& Kingham 1998), or BL~Lac (Denn, Mutel, \&
Marscher 1999) were detected to emerge.

The third catalog of high-energy (E$>$100~MeV) $\gamma$-ray sources detected by the
EGRET telescope on the Compton Gamma Ray Observatory  includes 66
high-confidence identifications of blazars and presents $\gamma$-ray light curves
from April 1991 to October 1995 (Hartman et al. 1999). 
Our sample of blazars 
observed with the VLBA contains 64\% of these $\gamma$-ray sources.
We  have detected more than 80 superluminal components ejected during 
the period from 1980 to 1996.5. The majority of (extrapolated) epochs of  zero separation
of components from the VLBI core fall into the interval from 1993 to 1996, since
the best sampling of our observations covers the  period
from October 1994 to November 1996. 
The components generally have high apparent speeds (the distribution
of apparent velocities peaks at 8-9~$h^{-1}c$, where
$H_\circ=100h$~km~s$^{-1}$~Mpc$^{-1}$, $q_\circ=0.1$) with a lifetime of visibility
of a typical component in the 22/43~GHz images of 
1.5$\pm$0.5~yr. Therefore, if every $\gamma$-ray flare is associated with the ejection
of a superluminal component, then $\gamma$-ray flares detected during the period
from 1993 to 1996 should correspond to VLBI components in our images. However,
we are not able to test the inverse relationship --- whether every ejected superluminal
component is associated with a $\gamma$-ray flare --- owing to the sparseness of the
$\gamma$-ray light curves and uncertainties of the extrapolated epochs of zero separation of
the superluminal components from the core. 
Typical uncertainties in the time of zero separation are $\sim$0.2~yr.
The characteristic time scale of a $\gamma$-ray outburst in blazars ranges from a
week to a month (Mattox et al. 1997b, McGlynn et al. 1997, Mattox et al. 2001), 
about 2 times less than the uncertainty in the time of component ejection. On average,
a $\gamma$-ray light curve consists of 16$\pm$5 measurements during 4.5~yr, 
half of which are only upper limits, usually when the sensitivity of
the observation is relatively low. Despite these
difficulties, we consider that our sample suffices to determine statistically
whether $\gamma$-ray
flares are associated with major energetic disturbances that propagate down the jet. 

\section{Gamma-Ray Flares} 

The identification of $\gamma$-ray flares is a key aspect of our statistical analysis,
since if there is a one-to-one correspondence between $\gamma$-ray flares and ejections
of superluminal components then every $\gamma$-ray flare detected during
the period from 1993 to 1996 should be related to the appearance of a superluminal feature 
in our images. From the entire database of $\gamma$-ray fluxes of
blazars collected in the 3rd EGRET catalog (Hartman et al. 1999) we have determined the average 
$\gamma$-ray flux of every source as a weighted mean 
of all measurements, including upper limits, 
with weight equal to $1/\sigma$, where $\sigma$ is the uncertainty of the measurement;
in the case of an upper limit $\sigma$ is equal to the value of the upper
limit itself. This definition yields
a large standard deviation of the mean compared with the average flux itself.
The average $\gamma$-ray flux is considered
as the base level of $\gamma$-ray emission, although it probably exceeds
the actual quiescent $\gamma$-ray state (if such a quiescent $\gamma$-ray state 
exists in blazars).  
We specify that a $\gamma$-ray flare occurred when a flux
measurement exceeded the mean flux by a factor of 1.5 or more and 
the uncertainty of the 
measurement at the high state was less than the deviation of the measurement from the 
mean. We make only one exception to these criteria: 
in the quasar 0528$+$134 we classify the $\gamma$-ray activity in the first
half of 1995 as a $\gamma$-ray flare (the third $\gamma$-ray flare for this source)
despite a ratio of the maximum measured flux
to the average flux equal to 1.4. In this case an increase in the $\gamma$-ray flux
was observed over 5 successive measurements (with small uncertainties) such that the structure of the flare is 
well defined (see Fig. 2a). Table 1 presents the results of $\gamma$-ray 
flare detections
for all blazars from our sample for which a high state of $\gamma$-ray
emission is found.  The columns of Table 1 are as follows: (1) 
source name, (2) average $\gamma$-ray flux in units of 
$10^{-8}$~phot~cm$^{-2}$~s$^{-1}$ ($<S_\gamma>$) and its standard deviation, (3) the ratio
of the maximum $\gamma$-ray flux during the flare to the average  
$\gamma$-ray flux ($f_\gamma$), (4) the epoch of the maximum 
$\gamma$-ray flux ($T_\gamma$), (5) an indication of 
whether the $\gamma$-ray flare is contemporaneous with the VLBA observations (Yes/No),
and (6) the number of superluminal ejections found during contemporaneous EGRET/VLBA
observations (1993--1996).
All together, we have identified 23 $\gamma$-ray flares (in 18 sources) that 
are contemporaneous with our VLBA observations.  
 
\section {Comparision of Epochs of Gamma-Ray Flares with Epochs
of Superluminal Ejections}

A comparison of epochs of $\gamma$-ray flares ($T_\gamma$) with epochs of zero separation
of superluminal components from the core ($T_\circ$) \footnote{The ``epoch of superluminal 
ejection'' is the time of coincidence of the position of the moving feature with the core
in the VLBA images, obtained from a linear least-squares fit to a plot of component-core 
separation vs. time. The entire list of 
epochs of superluminal ejections for all components detected during
our monitoring is given in Table 5 of Jorstad et al. 2001.} separates
the sample of $\gamma$-ray flares into three groups. Group $A$ includes
positive detections for which $|(T_\gamma-T_\circ)|\le\sigma(T_\circ)$,
where $\sigma(T_\circ)$ equals the 1$\sigma$ uncertainty of the epoch of
zero separation. The marginal detections represent group $B$, for which
$|(T_\gamma-T_\circ)|\le 3\sigma(T_\circ)$. Group $C$ (negative detections) 
contains those
$\gamma$-ray flares that do not appear to be related to ejections
of superluminal components according to the aforementioned criteria.

\subsection{Group $A$: Positive Detections}

In 10 of 23 $\gamma$-ray flares the epochs of zero separation
coincide, to within the 1$\sigma$ uncertainty, with the time of the
$\gamma$-ray flares. The list of positive detections includes
the sources 0336$-$019 (CTA~26), 0528$+$134, 0836$+$710, 1222$+$216,
1226$+$023 (3C~273), 1622$-$253, 1611$+$343 (DA 406), and 1730$-$130 (NRAO~530). 
For four of these
sources (0528$+$134, 0836$+$710, 1226$+$023, and 1611$+$343) a connection between $\gamma$-ray 
activity and ejections of
superluminal components was found previously based on different VLBI data 
(Britzen et al. 1999; Otterbein et al. 1998; Krichbaum et al. 1996; Piner \& Kingham 1998).
For quasar 1222$+$216 we have only 2 epochs of VLBA observations, and therefore 
cannot determine the uncertainty of the epoch of zero separation; 
we adopt an error of 0.2~yr.
In each of  the quasars 0528$+$134 and 1730$-$130 
two detected $\gamma$-ray flares were accompanied by superluminal ejections (see Fig. 2a).
In the case of 1730$-$130 the creation of both components appears to have been associated
with flares in radio light curves (Bower et al. 1997). 

\subsection{Group $B$: Marginal Detections}

The list of marginal detections includes 6 sources: 0420$-$014, 0440$-$003 (NRAO~190),
0458$-$020, 1219$+$285 (ON~231), 1253$-$055 (3C~279),  and 1622$-$297. 
In 1622$-$297 the ejection of a superluminal component occurred slighly earlier
than $T_\gamma-3\sigma(T_\circ)$ 
(note that in this case $\sigma(T_\circ)$ is smaller than usual). 
However, a very high level of $\gamma$-ray flux was observed one
week before the maximum $\gamma$-ray emission (Mattox et al. 1997b);
this indicates that the outburst was prolonged and started before $T_\gamma$. 

\subsection{Group $C$: Negative Detections}

The list of negative detections consists of 7 sources: 0827$+$243, 0917$+$449,
1101$+$384 (Mkn~421), 1127$-$145, 1219$+$285 (ON~231), 1222$+$216, and 1622$-$253. 
Since we are testing the simplest hypothesis that all $\gamma$-ray flares correspond 
to superluminal ejections, we examine whether there is a possible explanation for each
negative result in terms of observational uncertainties.
 
The last three of these sources each belongs to groups $A$ or $B$ as well,
i.e. more than one $\gamma$-ray flare was detected and at least one high state of 
$\gamma$-ray flux was accompanied by a superluminal ejection.  
In the  quasar 1222$+$216 the
$\gamma$-ray light curve contains two flares separated by an interval of one year;
in our two VLBA images we detected one component associated with the first 
$\gamma$-ray outburst (see Table 1) and found a second component with a speed
significantly faster than that of the first knot. This second component has an 
extrapolated time of zero separation from the core that is about two years later than
the time of the second $\gamma$-ray flare. However, the proper motions are estimates 
based on only two images, hence the uncertainties in the times of zero separation
could be much larger than our adopted value and the second component could be related
to the second $\gamma$-ray outburst.
Two of the three VLBA images of the BL Lac object 1219$+$285, obtained at closely
spaced epochs, show a significant change in the jet, but with such limited data we
can not identify components unambiguously across epochs; hence, two different 
interpretations of proper motions are possible (see Fig. 21b in Jorstad et al. 2001).
We adopt the lower proper motions,
similar to those obtained previously by Gabuzda et al. (1994). In this case
we find a connection between the $\gamma$-ray flare and a superluminal ejection that
occurred in the first half of 1994 (see Fig. 2a) but no association of a moving component
with the $\gamma$-ray outburst that took place about one year later. However, if we
consider the higher possible values of the proper motions
then contemporaneous VLBA data exist only for the second
$\gamma$-ray flare, for which a component emerged
from the core at 1995.1$\pm$0.2, within 1$\sigma$ uncertainty of the epoch
of the second high state of $\gamma$-ray emission. In 1622$-$253 the VLBA data show
one moving component while  the $\gamma$-ray light curve suggests two
flares. However, the detected maxima of the flares differ in time by less than 
0.3~yr. Although there are two measurements between these maxima
yielding  upper limits to the $\gamma$-ray emission similar to the
average $\gamma$-ray flux, the flares might be part of a single, complex, energetic
event.

Three 22~GHz images of 1127$-$145 show a weak component merging with 
a bright stationary feature 4~mas from the core, with the moving knot
appearing to decelerate during the last 
two epochs. Although our estimate of the time of zero separation is within 3$\sigma$
uncertainty of the epoch of the $\gamma$-ray flare, the apparent deceleration
leads to a very large (1.4~yr) uncertainty for the former. Hence, we do not 
classify this case as a marginal correspondence despite its meeting our formal 
criterion (see Fig 2b).  In Mkn~421
the jet components contain only a few percent of the total flux density and
the jet appears approximately the same from one epoch to the next. 
It is therefore very difficult to determine the apparent speed of features in the jet
and it is possible only to suggest the most likely interpetation. We have identified
components in the way which produces  apparent speeds $\sim$1.3--2~$h^{-1}c$
(Marscher 1999), while Piner et al. (2000)
interpret the data in terms of only subluminal motion. In the latter case our VLBA
observations would have taken place too early to resolve any component ejected
during the $\gamma$-ray flare.

In the quasars 0827$+$243 and 0917$+$449 we have sufficient VLBA data 
to reveal the motion of components present in the jet. 
However, in the case of 0827$+$243 the superluminal ejection nearest
to the high state of the $\gamma$-ray flux occurred about 0.5~yr
later than the $\gamma$-ray flare, with an uncertainty in 
the time of zero separaton of less than 0.1~yr.  In 0917$+$449
one moving component is detected but appears to have been ejected significantly
earlier than the time of maximum $\gamma$-ray flux. Therefore,
in these two cases we do not find any connection between the high state
of $\gamma$-ray emission and jet activity. 

\subsection{Statistical Simulations} 

We have carried out numerical simulations in order 
to determine the probability of 
random coincidences between epochs of $\gamma$-ray flares
and ejection of superluminal jet components.
To do this, we generated 1,000,000 samples of 18 sources (the number of objects is
equal to the number of blazars with contemporaneous EGRET/VLBA data)
with $\gamma$-ray flare/superluminal ejection features identical to our
observations. This means that for every source the number and epochs of $\gamma$-ray
flares were set according to Table 1 and random epochs of superluminal ejections
of VLBI components over the period of contemporaneous VLBA and $\gamma$-ray 
observations (1993-1996)
were generated in a quantity corresponding to the number of
superluminal ejections found during this period of observations (see Table 1). 
Therefore, each sample of 1,000,000 generated ones
consisted of 18 sources with 31 random superluminal ejections to compare with
23 $\gamma$-ray flares with fixed epochs. 
A coincidence was registered in the same manner as for groups $A$ and $B$  of
$\gamma$-ray bright blazars discussed above (1$\sigma$ and 3$\sigma$ criteria). 
Uncertainties of generated epochs of zero-separations are equal to the
uncertainties of epochs of {\it observed} superluminal ejections; in the case of
multiple ejections in a source the uncertainty of the first generated epoch equals
the uncertainty of the earliest superluminal ejection for the source, the uncertainty of the second
generated epoch is the same as for the next {\it observed} ejection and so on. 

Fig. 1 presents histograms of the probability density functions of chance coincidences 
for 1$\sigma$ (intense shade) and 3$\sigma$ (light shade) registration
criteria. (Note that the histograms do not overlap.) 
In the case of the 1$\sigma$ criterion the number of 9 chance coincidences
has a probability less than 1.5\%,
and in no case out of five independent runs of the 1,000,000 simulation samples
did chance coincidences occur for 10 of the sources in a sample. Therefore,
our finding of 10 positive detections out of 23 $\gamma$-ray flares
leads us to conclude that
high $\gamma$-ray states and superluminal ejections are associated
at $>$99.999\% confidence level. This does not necessarily imply that all 
$\gamma$-ray flares and superluminal ejections in group $A$ are physically 
associated: an average of 5--6 such pairings is expected by chance given
the uncertainties, especially, for uncertainties $\ge$0.5~yr.              
In the case of the 3$\sigma$ criterion the number of expected chance coincidences
falls in the range from 12 to 17 events  with a probability $>$92\%; and
the number of chance coincidences $\ge$16 (the number of coincidences which
we have detected for our sample of $\gamma$-ray bright blazars with 3$\sigma$ criterion) 
occurs with $\sim$21\% probability, too high to provide any meaningful constraints.

\section{Discussion}

Correlations between the emergence of new VLBI components and radio
flares have been noted by many authors (e.g., Mutel, Denn, \& Dryer 1994;
Wagner et al. 1995; Wehrle et al. 1998). Furthermore, polarized intensity VLBI images
of blazars reveal that superluminal knots are substantially polarized
(Gabuzda \& Cawthorne 1996; Lister, Marscher, \& Gear 1998; G\'omez,
Marscher, \& Alberdi 1999). Therefore, one should expect a 
correlation between the emergence of new VLBI components
and flares in polarized radio flux and hence between 
high $\gamma$-ray states and increases in polarized radio flux if $\gamma$-ray
flares are associated with superluminal ejections.

The University of Michigan Radio Astronomy Observatory database
allows us to test for these expected correlations. Table 2 presents the 
parameters of the radio polarization
for the positive and marginal $\gamma$-ray/superluminal ejection associations (Part I) 
as well as for the other $\gamma$-ray flares from Table 1
(negative detections or VLBA data that are not contemporaneous)
for which the UMRAO database contains sufficient polarization data (Part II). 
The columns of Table 2 are as follows: (1) the source name; (2) the epoch of the $\gamma$-ray
flare ($T_\gamma$); (3) the time of zero separation of the superluminal
component from the core with 1$\sigma$ uncertainty ($T_\circ$); (4) the frequency 
of the polarized radio flux density data ($\nu$); (5) the maximum polarized flux density nearest 
to the time of the $\gamma$-ray flare ($S_p^{max}$); (6) the epoch of 
the local maximum of polarized flux density ($T_p^{max}$);
(7) the nearest minimum polarized flux 
density before the local maximum of polarized flux density ($S_p^{min1}$); (8) 
the epoch of the first minimum of polarized flux density ($T_p^{min1}$); 
(9) the nearest minimum polarized flux density
after the local maximum of polarized flux density ($S_p^{min2}$); and (10) 
the epoch of the second  minimum of polarized flux density ($T_p^{min2}$). 
Fig. 2a presents $\gamma$-ray light curves and total as well as
polarized radio light curves, along with indications of the epochs of the maxima
in $\gamma$-ray flux
and superluminal ejections for the positive and marginal detections (Groups $A$ and $B$),
and Fig. 2b illustrates several examples for cases in Part II of Table 2.
Although many of the polarized flux density light curves are noisy, 
it is apparent that, shortly after a superluminal
ejection and $\gamma$-ray flare, a local maximum in polarized radio flux is observed. 

In order to relate the multifrequency data, we have calculated the
differences between the epoch of each $\gamma$-ray flare and the time of 
superluminal ejection as well as maximum and minimum level of polarized
flux density ($T_\circ-T_\gamma$, 
$T^{max}_p-T_\gamma$, $T^{min1}_p-T_\gamma$, $T^{min2}_p-T_\gamma$). We then normalized
the polarized flux densities in terms of the local
maximum level for every case of positive and marginal associations of
$\gamma$-ray flares and superluminal ejections (Groups $A$ and $B$). 
Finally, we averaged these values over the entire sample. The result 
is presented in Fig. 3a, which reveals the general pattern of
behavior. A disturbance causes the appearance of a superluminal component
near the time of minimum polarized flux density
($<T^{min1}_p-T_\circ>=-17\pm68$~days). A $\gamma$-ray flare follows
$52\pm76$~days later, occurring almost simultaneously
with the local maximum of the polarized radio flux density ($<T^{max}_p-T_\gamma>=7\pm38$~days).
During the outburst the polarized flux density nearly doubles 
($<S^{min1}_p/S^{max}_p>=0.42\pm0.23$) and returns to the initial quiescent state
($<S^{min2}_p/S^{max}_p>=0.38\pm0.21$) 98$\pm$66~days after the $\gamma$-ray flare.
The entire duration of the combined event is 170$\pm$135~days.

Similar calculations were performed for cases listed in Part II of Table 2,
where no connection between $\gamma$-ray flares and superluminal
ejections could be established but variability of the radio polarized flux density was observed 
(Fig. 3b).
The scenario of the perturbation is very similar to that decribed above. The duration
of the event is 172$\pm$110~days, the first minimum of polarized flux density occurs
30$\pm$52~days before a $\gamma$-ray flare, and the maximum polarized flux density occurs close
to the time of the maximum $\gamma$-ray emission, with a delay of only 15$\pm$20~days. 
The polarized flux
density returns to the quiescent state 142$\pm$57~days after the $\gamma$-ray flare.      
  
The results given in Table 2 support the idea 
that the radio emission and the $\gamma$-rays
originate within the same shocked region of the relativistic jet and that the $\gamma$-rays
are most likely produced by inverse Compton scattering by the electrons in the jet {\it far
(typically parsecs) downstream of the putative accretion disk}.
The generalized diagrams (Fig. 3) lead us to propose a speculative model to
explain the derived characteristics of such an event. 
An emerging jet component is cross-polarized with respect to the polarization
of the VLBI core (e.g., in the quasar 3C~454.3; Gomez, Marscher, \&
Alberdi 1999); this results in a decrease of the total polarized flux density ({\it the first minimum in polarized flux 
density}) during the birth of the polarized component. Shortly after the local minimum of polarized flux density a superluminal
ejection occurs.  The standard model explains the VLBI component as a manifestation of a
shock wave propagating through an underlying relativistic outflow (e.g.,
Marscher \& Gear 1985).  
The shock front orders the magnetic field, hence the polarization
increases when the front first becomes optically thin ({\it the maximum in polarized
flux density}). (Evidence for this is found in the observed increase in the fractional linear
polarization of four components in BL~Lac as each knot separated
from the core;  Denn, Mutel, \& Marscher 2000.) The $\gamma$-ray emission is caused
by inverse Compton scattering in the thin forward layer 
of the shock, where electrons are accelerated. This high-energy emission is first detected 
somewhat downstream of the radio core.

The average delay of 52 days between the
extrapolated time $T_\circ$ when the superluminal knot (which corresponds to the shocked
plasma) is coincident with the radio core and the epoch of the $\gamma$-ray flare
could correspond to the time required for (1) the shock to develop fully in terms
of efficiency of accelerating electrons and compressing the magnetic field, (2) the bulk
velocity of the shock to accelerate to its asymptotic value, (3) the Doppler factor
of the shock to reach a maximum owing to the
changing angle between the velocity vector and the line of sight as it travels through a 
bend in the jet, or (4) the synchrotron
photons to travel across the shock to the scattering electrons (synchrotron
self-Compton model only; see Marscher 2001). The large scatter in the difference
between $T_\circ$ and $T_\gamma$ indicates that there is significant variation
among blazars, as one would expect to be the case at least for explanations (3)
and (4) that depend sensitively on the angle between the jet axis and the line of sight.
The near-coincidence between the peak in $\gamma$-ray flux and polarized radio flux
density can most easily be accommodated by scenarios (1)--(3), since both the
synchrotron and inverse Compton emissivities of the shock would vary together.
The absence of a clear signature in the total radio flux density light curve could
be caused by the core fading as the superluminal knot brightens, for example under 
the influence of a rerefaction in the wake of the shock (cf. G\'omez et al. 1997).

The distance traveled by the disturbance between the zero-separation time $T_\circ$
and the peak in the $\gamma$-ray flux and polarized radio flux density, typically
$\Delta t \sim 2$ months, is given by
$\Delta r \approx \Gamma \delta c (1+z)^{-1} \Delta t$, where $\Gamma$ and $\delta$
are, respectively, the bulk Lorentz factor and the Doppler factor of the disturbance.
For typical values of apparent superluminal velocities of $\gamma$-ray blazars $\sim 12~c$
found by Jorstad et al. (2001) (for $h = 0.65$), this corresponds to several parsecs.
If the jet opening angle is $\sim 1^\circ$, the transverse size of the emission
region at this distance is of order 0.1 pc. The shortest timescale of variability
should therefore be $\sim 100\zeta (1+z)/\delta$ days, or typically $\sim 10 \zeta$
days, where $\zeta \leq 1$ is a geometrical factor that allows for a line-of-sight
thickness of the shock that is much less than the size in the sky plane (Marscher \& Gear 1985).
Under this scenario, the most rapid variations in the $\gamma$-ray flux
observed by EGRET, on scales of $\le 1$ day (e.g., Mattox et al. 1997b;
Wehrle et al. 1998), represent extreme cases of high
Doppler factors (the data of Jorstad et al. 2001 indicate values exceeding 30
in a number of sources) and/or values of $\zeta$ that are significantly less than
unity.

In the transverse shock model we should expect the magnetic field to become perpendicular
to the jet axis when the front of the shock becomes optically thin
({\it the maximum in polarized flux density}). However, Lister, Marscher, \& Gear (1998)
found that the majority of the components in blazar jets 
observed at 43~GHz (sample of 8 sources) have oblique orientations of their electric vectors. 
Furthermore, Denn, Mutel, \& Marscher (1999) reported no
preferred orientation of the initial position angle of the electric vectors of
superluminal components in BL~Lac. Rather, they found a trend toward parallel alignment 
(perpendicular magnetic field) as components moved away 
from the core. To check whether similar behavior occurred in our sample, we list
in Table~3 position angles of polarization for the sources from Table~2 for which
polarization data are available with at least 1$\sigma$ determinations of the
position angle of the electric vector.
The columns of Table~3 are:
(1) source name; (2) position angle (PA) of the jet on submilliarcsecond  
scales ($\Theta_{jet}$); (3) frequency of the polarization data ($\nu$);
(4) position angle of the electric vector (EVPA) at the time of maximum polarized
flux density ($\phi_p^{max}$); (5) EVPA at the first minimum in polarized flux density
($\phi_p^{min1}$); (6) EVPA at the second minimum polarized flux density ($\phi_p^{min2}$);
(7) orientation of the magnetic field with respect to the jet direction at the
maximum polarized flux density ($\perp$: transverse --- difference between PA of
the jet and EVPA $<$20$^\circ$; $\parallel$: parallel --- difference between
PA of the jet and EVPA $>$70$^\circ$; and $<$: oblique --- all other cases).
No Faraday rotation measure (RM) corrections are applied to the observed EVPAs, since even
the relative large RM for quasar 0458$-$020 ($-$170~rad~m$^{-2}$; Simard-Normandin, 
Kronberg, \& Button 1981) produces a small Faraday rotation ($\le$5$^\circ$) at 15~GHz.
Therefore, we take the orientation of the magnetic field to be essentially perpendicular 
to the observed EVPAs.
Analysis of the data in Table 3 shows that in 9 out of 22 cases the
magnetic field had an oblique orientation at the maximum polarized flux density; however,
in the quasars 0458$-$020 and 1510$-$089 the magnetic field tended toward the transverse direction
at the maximum compared with the first minimum of
polarized flux density (when the magnetic field was also oblique). 
In 5 of 22 cases the magnetic field was almost perpendicular to
the jet direction at the high state of polarized flux density and  in the quasars  
0440$-$003 and 1611$+$343 a significant change of EVPA between the maximum and second minimum
of polarized flux density was observed. The remaining 8 cases
showed parallel orientation of the magnetic field at both the minima and 
maximum in polarized flux density; this implies a strong underlying parallel magnetic field.
These magnetic field orientations during the event are considerably more
complex than one would expect from the transverse or moving oblique shock models:
for the latter model Lister, Marscher, \& Gear (1998) showed using numerical
simulations that the distribution of EVPA misalignment angles should peak at
low misalignment values. The cause of this discrepancy between theory and observation
could be the overly simple nature of the transverse shock model. Numerical
hydrodynamical simulations (e.g., G\'omez et al. 1997) show that the structure of
a relativistic shock is likely to be quite complex, including secondary shocks and
trailing shocks and rarefactions.

\section{Conclusion} 

Our study establishes a statistical association between the ejection   
of superluminal radio knots and high states of $\gamma$-ray emission in 
blazars. Our analysis of the delay between the time when a superluminal
radio knot is coincident with the radio core and the epoch of the maximum
observed $\gamma$-ray flux, along with an investigation of the variability
of polarized radio flux density during the $\gamma$-ray flare, lead us to conclude 
that the radio and $\gamma$-ray events both originate within the same shocked area 
of the relativistic jet. This region is downstream of the core rather than between
the core and the central engine, as had been supposed previously, consistent
with the finding of  Valtaoja \& Ter\"asranta (1996)
that $\gamma$-ray flares occur after the mm-wave flux starts to rise.
The placing of the $\gamma$-ray emitting region downstream of the radio core
strongly supports inverse Compton models for the origin of the $\gamma$-rays.
The electrons accelerated in the shocks (which also correspond to superluminal
radio knots) scatter photons that originate either within the shocked region
or from a site external to the jet (e.g., the emission-line clouds; Sikora, Begelman, \& Rees
1994). Our results, on the other hand, conflict with models in which the
$\gamma$-rays are produced in the intense radiation environment close to the
putative accretion disk.

The relations between the times of radio and $\gamma$-ray events presented here
are rather rough owing to incomplete time coverage in both the $\gamma$-ray
light curves and the VLBA observations. Our statistical associations therefore
should be verified through
better time sampling of the $\gamma$-ray light curves of more blazars, coupled with
regular, closely spaced VLBA observations. The authors hope that a concerted
effort can be carried out with the VLBA when the planned GLAST $\gamma$-ray mission
provides well-sampled monitoring of the $\gamma$-ray fluxes of hundreds of blazars.

\begin{acknowledgments}

This work was supported in part by NASA through CGRO Guest Investigator
grants NAG5-7323 and NAG5-2508, and by U.S. National Science Foundation
grant AST-9802941. A.E.W. and J.R.M were supported in part by the NASA 
Long Term Space Astrophysics program. 
The UMRAO observations were partially supported by NSF grant AST-9421979 and 
preceeding grants and by the University of Michigan.
We are thankful to J. Ulvestad for substantional comments about the statistical
simulations.
\end{acknowledgments}

\begin{deluxetable}{lccccc}
\singlespace
\tablenum 1
\tablecolumns{6}
\tablecaption{\small\bf Detections of Gamma-Ray Flares}
\tabletypesize{\footnotesize}
\tablehead{
\colhead{Source}&\colhead{$<S_\gamma>$}&\colhead{$f_\gamma$}&\colhead{$T_\gamma$}&\colhead{Contemporaneous}
&\colhead{Number of}\\ 
\colhead{}&\colhead{[$10^{-8}$phot cm$^{-2}$ s$^{-1}$]}&\colhead{}&\colhead{}&\colhead{VLBA obs.?}
&\colhead{Ejections} \\
}
\startdata
0202$+$149&21$\pm$20&2.5&1992.315&No&\nodata \\
0235$+$164&30$\pm$20&2.2&1992.185&No&\nodata \\
0336$-$019&34$\pm$74&5.3&1995.266&Yes&2 \\
0420$-$014&27$\pm$23&1.9&1992.185&No&1 \\
&&2.4&1995.622&Yes \\
0440$-$003&28$\pm$26&3.1&1994.632&Yes&1   \\
0458$-$020&21$\pm$21&3.2&1994.212&Yes&1  \\
0528$+$134&89$\pm$78&1.7&1991.315&No&4   \\
&&3.9&1993.233&Yes \\
&&1.4&1995.373&Yes \\
0716$+$714&20$\pm$12&2.3&1992.197&No&\nodata  \\
0827$+$243&37$\pm$36&3.0&1994.160&Yes&1\\
0836$+$710&18$\pm$9&1.9&1992.197&Yes&1 \\
0917$+$449&21$\pm$9&1.9&1994.190&Yes&1   \\
1101$+$384&17$\pm$7&1.6&1994.367&Yes&2  \\
1127$-$145&27$\pm$17&2.3&1993.024&Yes&1 \\
1156$+$295&27$\pm$26&6.1&1993.024&No&\nodata \\
1219$+$285&21$\pm$17&1.6&1992.984&No&3 \\
&&1.5&1994.367&Yes \\
&&2.6&1995.334&Yes \\
1222$+$216&21$\pm$14&2.3&1992.984&Yes&1  \\
&&2.1&1993.964&Yes\\ 
1226$+$023&22$\pm$14&2.2&1993.899&Yes&2 \\
1253$-$055&52$\pm$48&5.1&1991.474&No&3 \\
&&2.0&1993.879&Yes \\
1406$-$076&40$\pm$36&3.2&1993.008&No   \\
1510$-$089&31$\pm$12&1.6&1992.260&No \\
1606$+$106&35$\pm$16&1.8&1991.964&No \\
1611$+$343&36$\pm$23&1.9&1994.841&Yes&1  \\
1622$-$253&41$\pm$21&1.9&1995.460&Yes&1   \\*
&&2.0&1995.729&Yes   \\
1622$-$297&68$\pm$99&4.7&1995.460&Yes&3  \\
1633$+$382&53$\pm$33&2.0&1991.707&No&\nodata \\
1730$-$130&48$\pm$30&1.7&1994.553&Yes&2 \\*
&&2.2&1995.510&Yes \\
2230$+$114&30$\pm$13&1.7&1992.647&No&\nodata \\
2251$+$158&52$\pm$29&2.2&1992.647&No&\nodata \\
\enddata
\end{deluxetable}

\begin{deluxetable}{clllcllllll}
\singlespace
\tablenum 2
\tablecolumns{11}
\tablecaption{\small\bf Epochs of Superluminal Ejections and $\gamma$-Ray and  
Polarized Radio Flux Peaks}
\tabletypesize{\scriptsize}
\tablehead{
\colhead{Part}&\colhead{Source}&\colhead{$T_\gamma$}&\colhead{$T_o$}&\colhead{$\nu_p$}&\colhead{$S_p^{max}$}& 
\colhead{$T_p^{max}$}&\colhead{$S_p^{min1}$}&\colhead{$T_p^{min1}$}&
\colhead{$S_p^{min2}$}&\colhead{$T_p^{min2}$} \\ 
\colhead{}&\colhead{}&\colhead{}&\colhead{}&\colhead{[GHz]}&\colhead{[mJy]}&
\colhead{}&\colhead{[mJy]}&\colhead{}&\colhead{[mJy]}&
\colhead{} 
}
\startdata
I&0336$-$019&1995.266&1995.1$\pm$0.2&8.0&137$\pm$60&1995.361&31$\pm$16&1995.131&29$\pm$27$^*$&1995.415 \\
&0420$-$014&1995.622&1995.3$\pm$0.1&14.5&208$\pm$18&1995.432&75$\pm$16&1994.989&94$\pm$16&1995.760 \\
&0440$-$003&1994.632&1994.2$\pm$0.2&14.5&202$\pm$101&1994.740&9$\pm$13$^*$&1994.656&18$\pm$13$^*$&1994.828   \\
&0458$-$020&1994.212&1994.0$\pm$0.1&14.5&89$\pm$49&1994.333&30$\pm$11&1994.033&23$\pm$13&1994.705  \\
&0528$+$134&1993.233&1993.2$\pm$1.6&4.8&29$\pm$16&1993.257&12$\pm$5&1992.721&5$\pm$7$^*$&1993.926  \\
&&1995.373&1995.5$\pm$0.1&4.8&26$\pm$11&1995.213&6$\pm$6$^*$&1995.145&10$\pm$7&1995.792  \\
&0836$+$710&1992.197&1992.1$\pm$0.3&14.5&138$\pm$24&1992.203&117$\pm$16&1992.034&48$\pm$15&1992.274 \\
&1219$+$285&1994.367&1994.0$\pm$0.2&14.5&79$\pm$14&1994.414&24$\pm$13&1994.346&16$\pm$17&1994.706 \\
&1222$+$216&1992.984&1993.1$\pm$0.2& no data \\ 
&1226$+$023&1993.899&1993.7$\pm$0.3&14.5&1810$\pm$106&1994.031&1249$\pm$108&1993.837&1379$\pm$63&1994.165 \\
&1253$-$055&1993.879&1993.5$\pm$0.2&14.5&383$\pm$56&1993.995&132$\pm$38&1993.837&272$\pm$59&1994.031 \\
&1611$+$343&1994.801&1995.3$\pm$1.4&14.5&123$\pm$21&1994.854&53$\pm$12&1994.209&34$\pm$21&1995.073  \\
&1622$-$253&1995.729&1996.0$\pm$0.3& no data \\
&1622$-$297&1995.460&1995.1$\pm$0.1& no data  \\
&1730$-$130&1994.553&1994.6$\pm$0.2&8.0&297$\pm$27&1994.575&146$\pm$16&1994.382&161$\pm$85&1994.605 \\
&&1995.510&1995.5$\pm$0.1&14.5&356$\pm$19&1995.431&264$\pm$19&1995.390&175$\pm$21&1995.742 \\
II&0202$+$149&1992.315&\nodata&8.0&140$\pm$23&1992.486&29$\pm$33$^*$&1992.399&13$\pm$18$^*$&1992.579 \\
&0235$+$164&1992.185&\nodata&14.5&51$\pm$19&1992.200&23$\pm$18&1992.104&20$\pm$27$^*$&1992.208 \\
&0716$+$714&1992.197&\nodata&14.5&50$\pm$17&1992.203&13$\pm$27$^*$&1991.732&8$\pm$24$^*$&1992.492  \\
&1101$+$384&1994.367&\nodata&14.5&51$\pm$17&1994.367&9$\pm$16$^*$&1994.345&4$\pm$19$^*$&1994.501  \\
&1127$-$145&1993.024&\nodata&8.0&134$\pm$27&1993.056&86$\pm$20&1992.913&89$\pm$19&1993.160 \\
&1156$+$295&1993.024&\nodata&14.5&83$\pm$16&1993.026&24$\pm$17&1993.000& 37$\pm$14&1993.542 \\
&1510$-$089&1992.260&\nodata&14.5&130$\pm$61&1992.272&68$\pm$16&1992.231&56$\pm$16&1992.373 \\
&1606$+$106&1991.964&\nodata&8.0&41$\pm$54$^*$&1992.051&36$\pm$19&1991.834&24$\pm$23$^*$&1992.157 \\
&2230$+$114&1992.647&\nodata&14.5&78$\pm$15&1992.666&29$\pm$13&1992.573&48$\pm$18&1992.745 \\
&2251$+$158&1992.647&\nodata&14.5&266$\pm$79&1992.748&151$\pm$20&1992.620&170$\pm$36&1992.805 \\
\enddata
\tablecomments{entries for $S_p$ denoted with ($*$) are the upper limits of polarized flux
density detections} 
\end{deluxetable}

\begin{deluxetable}{clrcrrrc}
\singlespace
\tablenum 3
\tablecolumns{8}
\tablecaption{\small\bf Comparison between Position Angle of Radio Jets and  
Position Angle of Total Linear Radio Polarization}
\tabletypesize{\footnotesize}
\tablehead{
\colhead{Part}&\colhead{Source}&\colhead{$\Theta_{jet}$}&\colhead{$\nu_p$}&
\colhead{$\phi_p^{max}$}&\colhead{$\phi_p^{min1}$}&
\colhead{$\phi_p^{min2}$}&\colhead{Orientation of} \\ 
\colhead{}&\colhead{}&\colhead{[$^\circ$]}&\colhead{[GHz]}&\colhead{[$^\circ$]}&
\colhead{[$^\circ$]}&\colhead{[$^\circ$]}&\colhead{Mag.Field}
}
\startdata
I&0336$-$019&45&8.0&149$\pm$7&141$\pm$22& \nodata  &$\parallel$ \\
&0420$-$014&$-$160&14.5&94$\pm$3&130$\pm$7&111$\pm$5&$\parallel$ \\
&0440$-$003&$-$123&14.5&72$\pm$12& \nodata &45$\pm$16&$\perp$  \\
&0458$-$020&$-$7&14.5&26$\pm$15&47$\pm$13&42$\pm$14&$<$  \\
&0528$+$134&61&4.8&111$\pm$13&124$\pm$11& \nodata &$<$ \\
&&&4.8&123$\pm$9& \nodata &113$\pm$17&$<$ \\
&0836$+$710&$-$140&14.5&111$\pm$4&104$\pm$3&112$\pm$8&$\parallel$ \\
&1219$+$285&102&14.5&63$\pm$4&54$\pm$13& \nodata &$<$ \\
&1226$+$023&$-$108&14.5&150$\pm$1&144$\pm$2&149$\pm$1&$\parallel$ \\
&1253$-$055&$-$115&14.5&69$\pm$4&76$\pm$7&63$\pm$6&$\perp$ \\
&1611$+$343&173&14.5&1$\pm$3&7$\pm$6&126$\pm$14&$\perp$ \\
&1730$-$130&17&8.0&80$\pm$2&80$\pm$3&87$\pm$7&$<$ \\
&&&14.5&93$\pm$1&103$\pm$2&84$\pm$3&$\parallel$ \\
II&0202$+$149&$-$32&8.0&74$\pm$4& \nodata & \nodata &$\parallel$ \\
&0235$+$164&0&14.5&123$\pm$11&24$\pm$22& \nodata &$<$ \\
&0716$+$714&10&14.5&24$\pm$9& \nodata & \nodata &$\perp$ \\
&1101$+$384&$-$25&14.5&129$\pm$9& \nodata & \nodata &$<$ \\
&1127$-$145&86&8.0&160$\pm$6&167$\pm$10&158$\pm$5&$\parallel$ \\
&1156$+$295&0&14.5&149$\pm$4&170$\pm$19&161$\pm$10&$<$ \\
&1510$-$089&$-$32&14.5&115$\pm$13&90$\pm$7&68$\pm$8&$<$ \\
&2230$+$114&132&14.5&115$\pm$5&123$\pm$14&107$\pm$10&$\perp$ \\
&2251$+$158&$-$66&14.5&31$\pm$8&24$\pm$3&17$\pm$6&$\parallel$ \\
\enddata
\end{deluxetable}

\clearpage
\begin{figure}
\figurenum{1}
\plotone{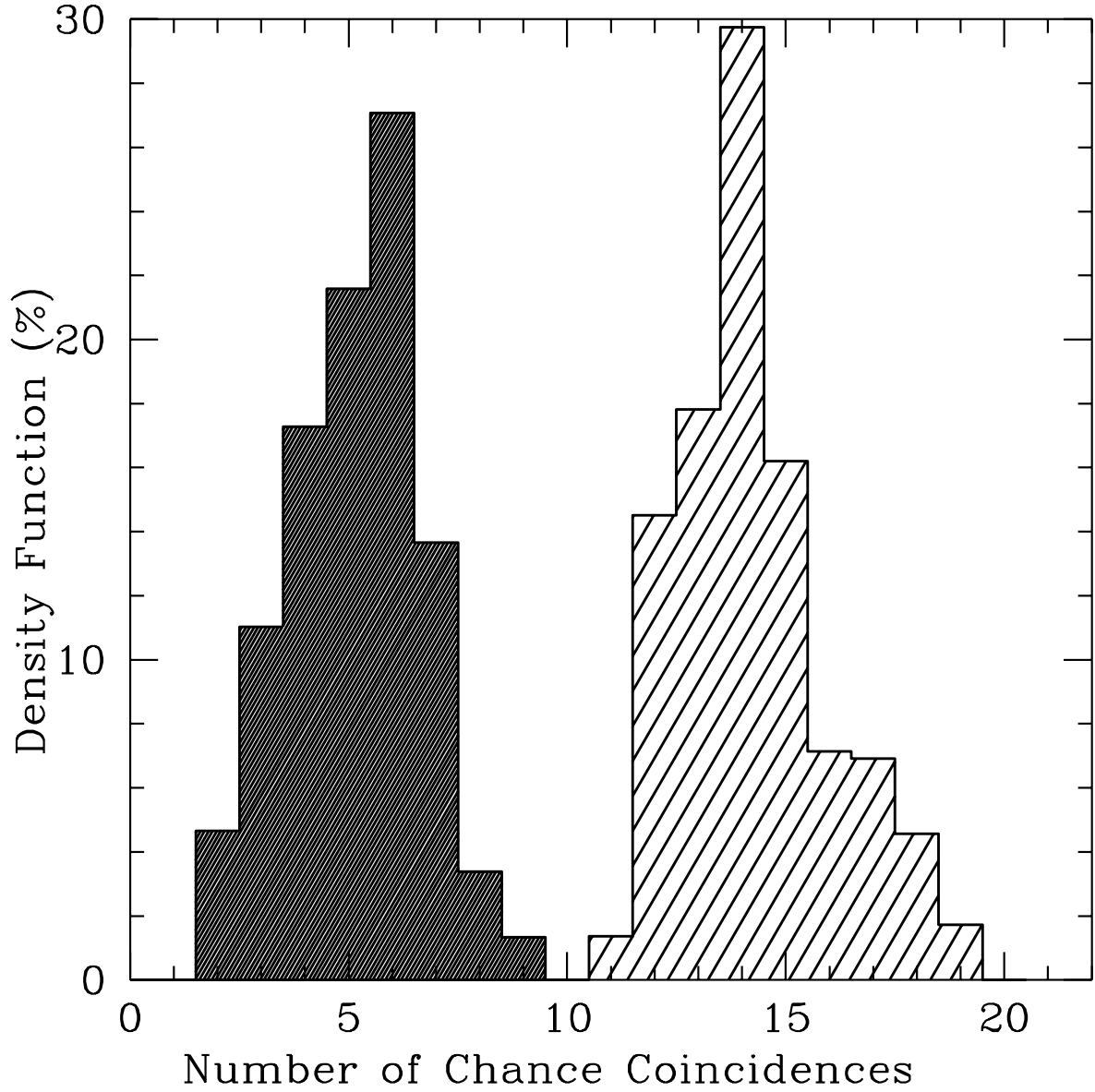}
\caption{Probability density functions of chance coincidences between
$\gamma$-ray flares and epochs of zero separation within 1$\sigma$
(intense shade) and 3$\sigma$ (light shade) uncertainties 
of epochs of superluminal ejections. }
\end{figure}
\begin{figure}
\figurenum{2a}
\plotone{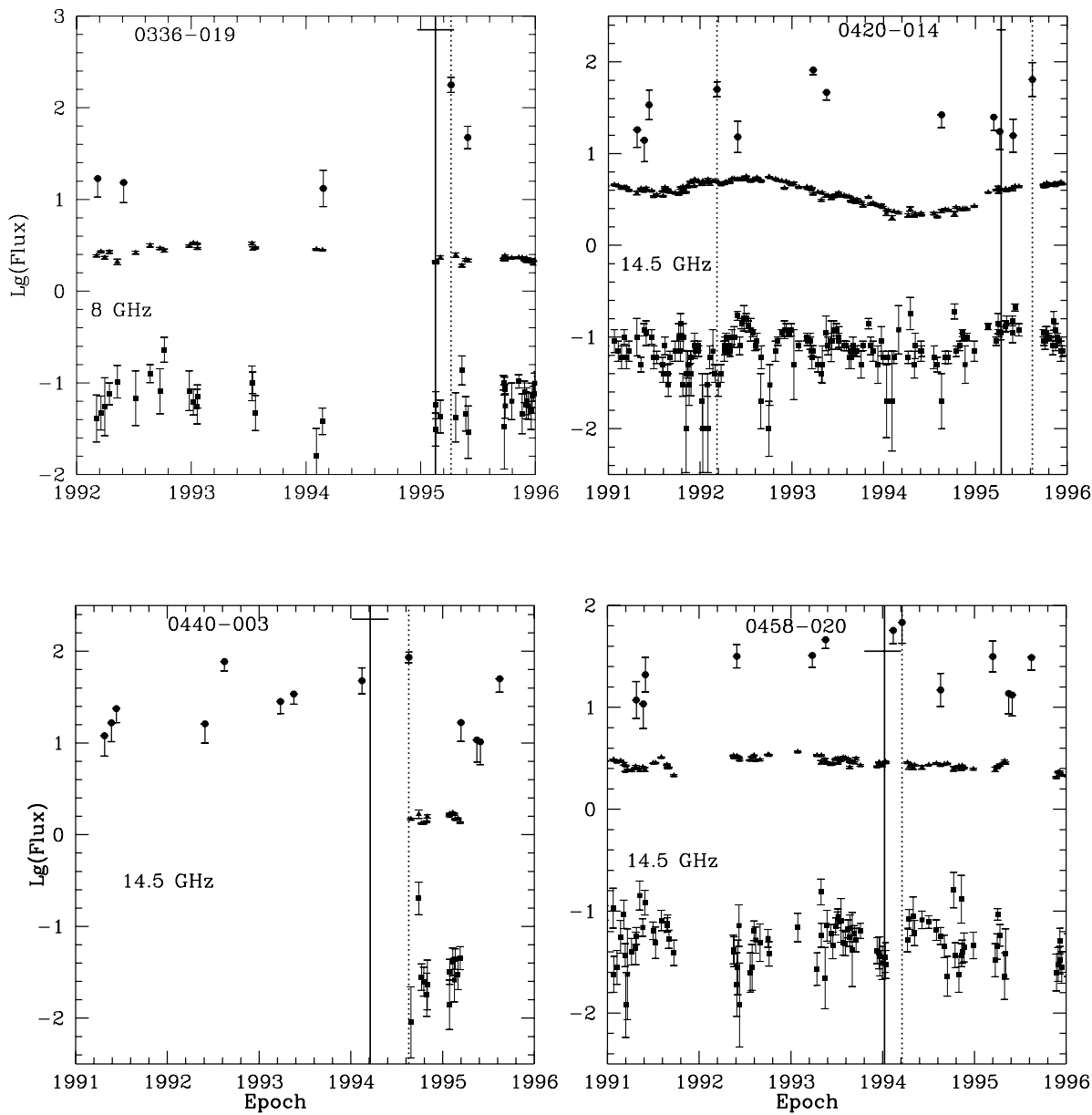}
\vspace{-4cm}
\caption{Radio and $\gamma$-ray data for the positive or marginal
coincidences between $\gamma$-ray flares and superluminal ejections:
$\gamma$-ray flux (circles), total radio flux density (triangles),
and polarized radio flux density (squares) on a logarithmic scale; 
solid lines indicate extrapolated times of zero
separation between radio knots and cores with 1$\sigma$ uncertainties, 
and dotted lines correspond 
to observed
maxima of the $\gamma$-ray emission.}
\end{figure}
\begin{figure}
\figurenum{2a}
\plotone{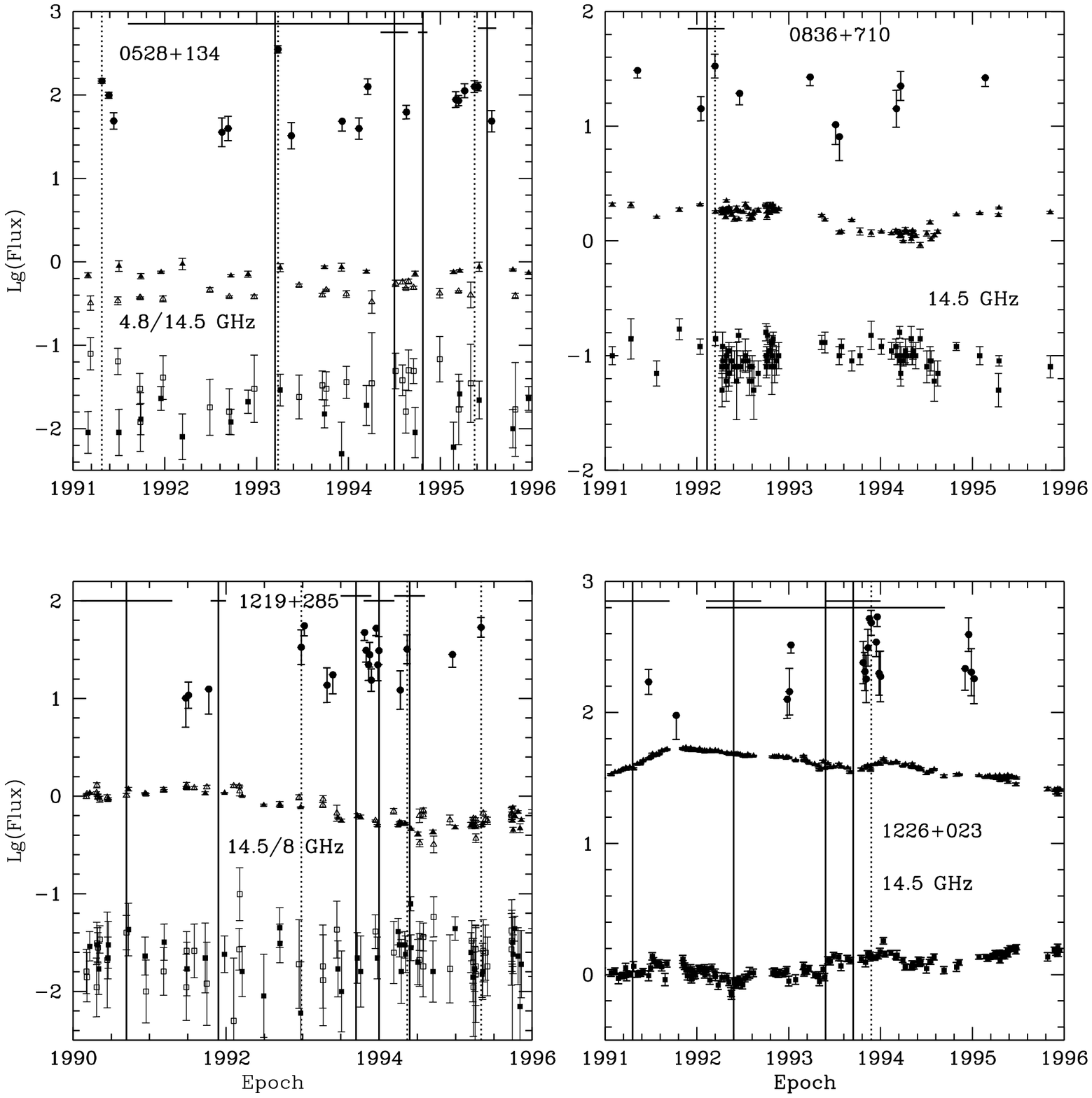}
\vspace{-4cm}
\caption{Continued}
\end{figure}
\begin{figure}
\figurenum{2a}
\plotone{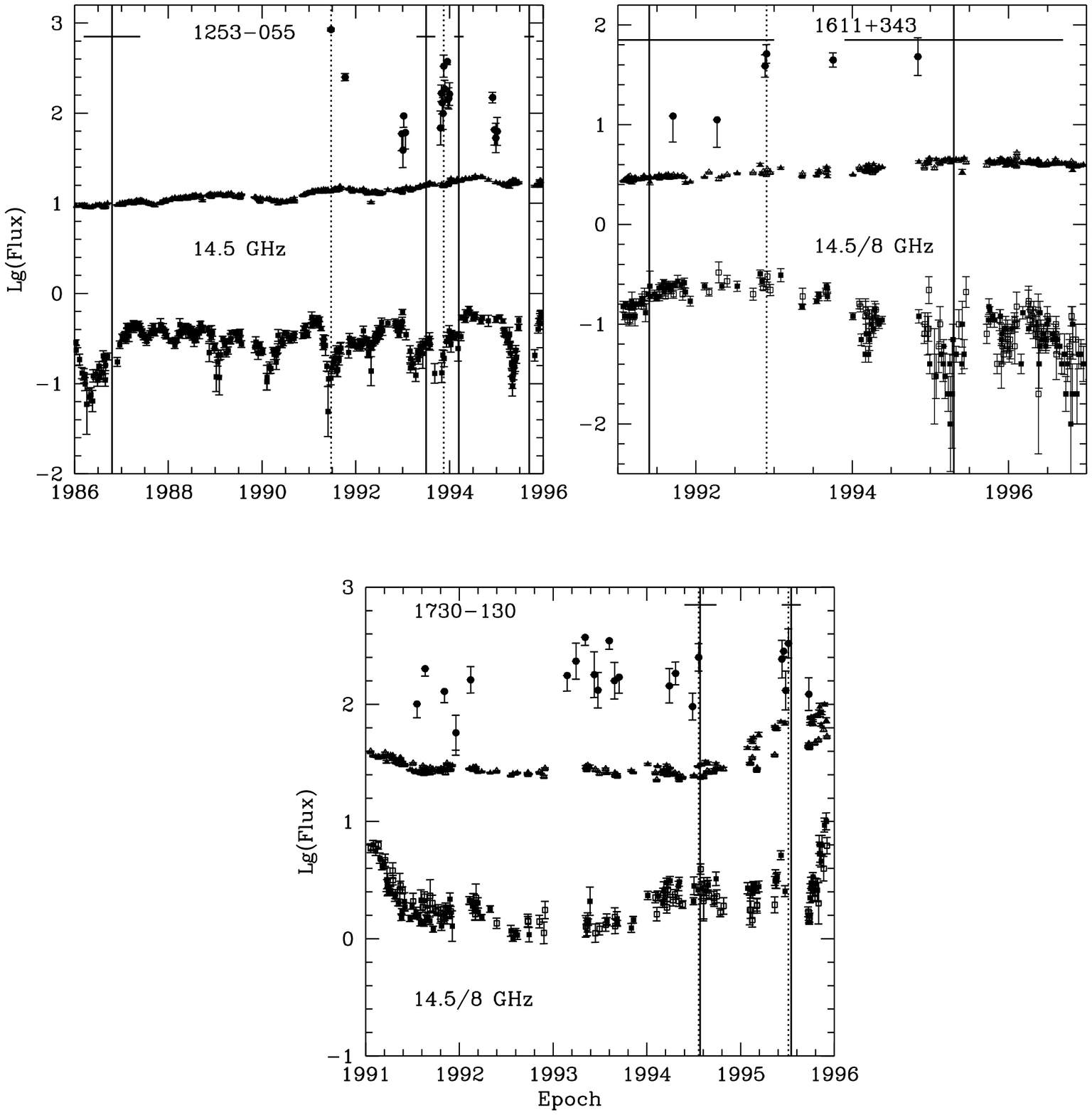}
\vspace{-4cm}
\caption{Continued}
\end{figure}
\begin{figure}
\figurenum{2b}
\plotone{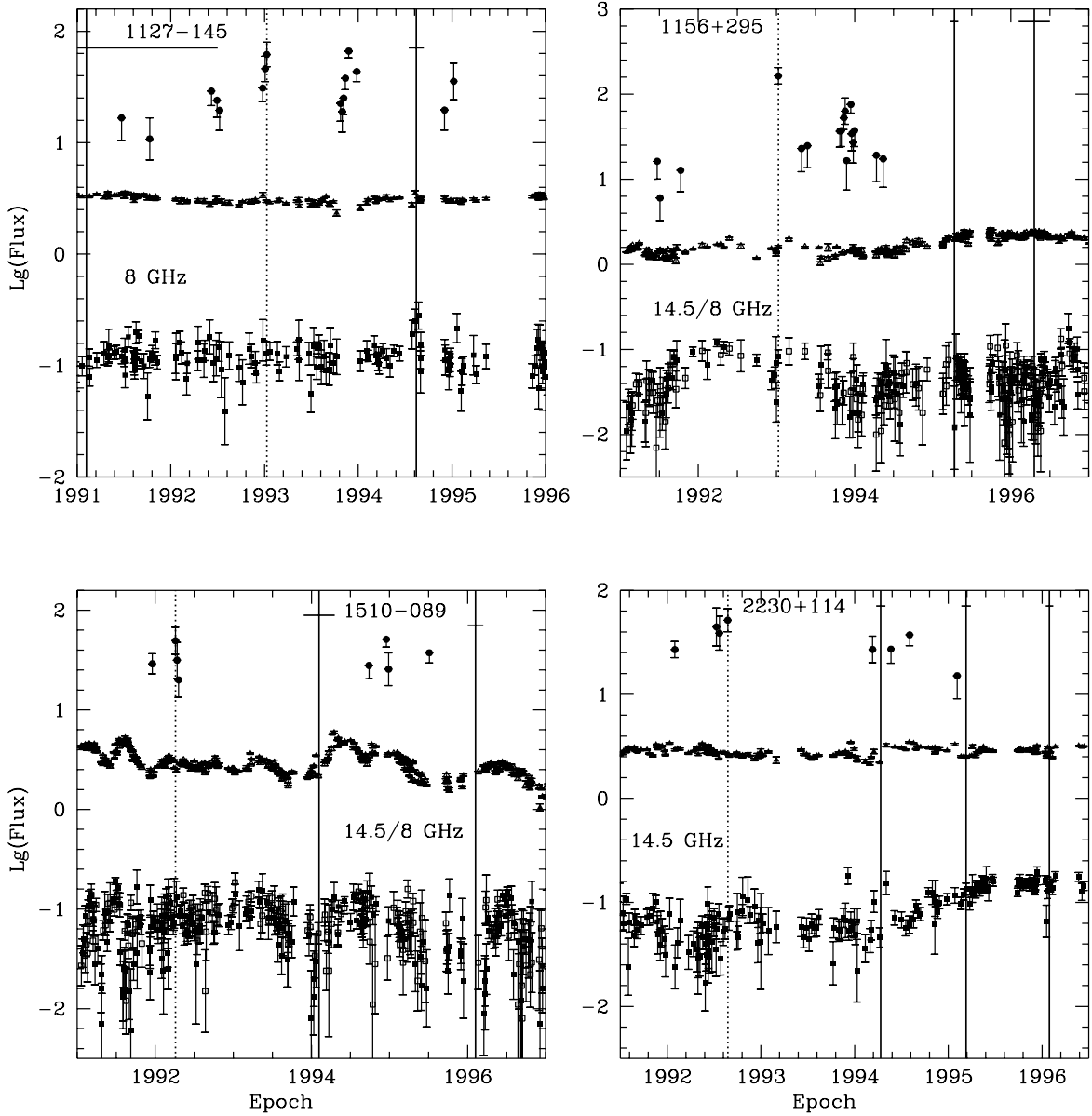}
\vspace{-4cm}
\caption{As in Fig. 2a but for $\gamma$-ray flares which are not accompanied 
by detected superluminal ejections.}
\end{figure}
\begin{figure}
\figurenum{3a}
\plotone{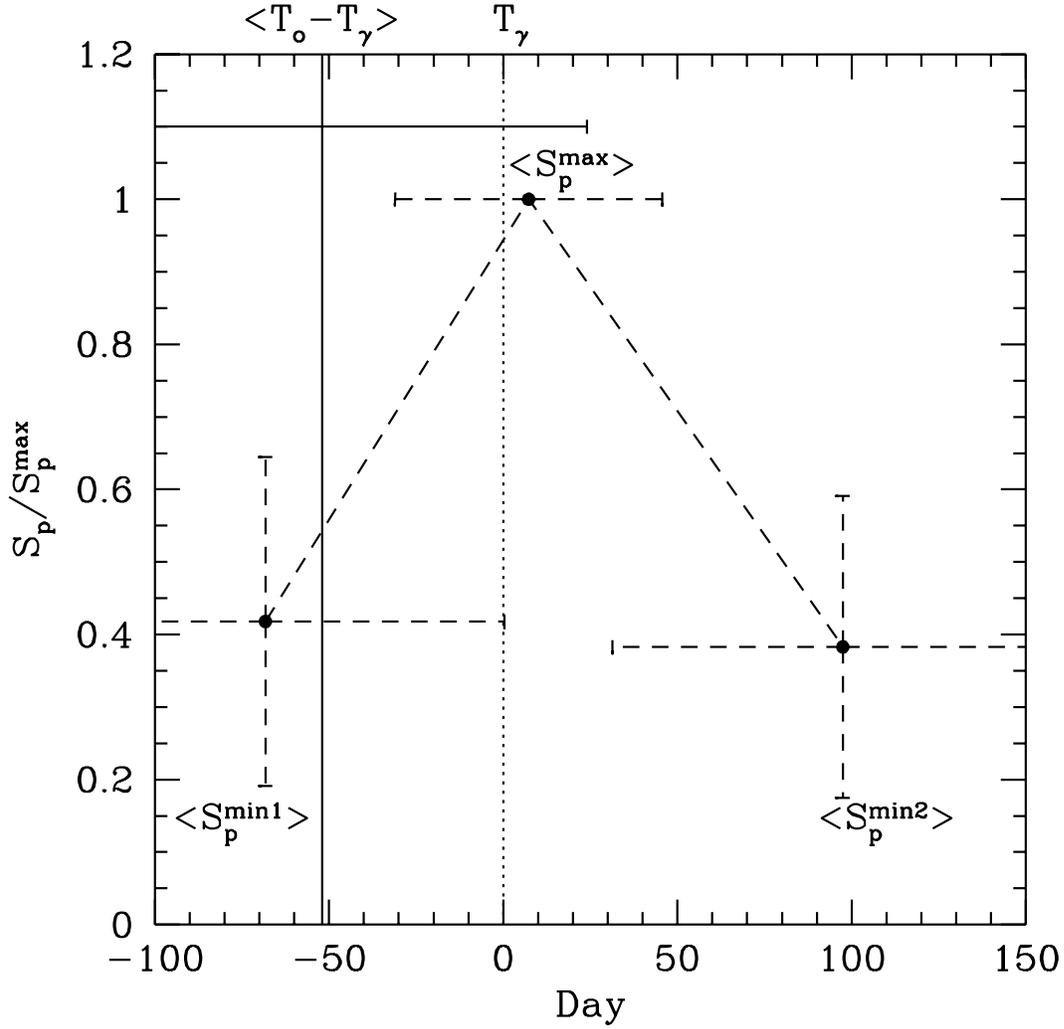}
\caption{Average temporal relationship between $\gamma$-ray flares,
superluminal ejections, and polarized radio flux densities. The dotted line indicates 
the epoch of the $\gamma$-ray flare, the
vertical solid line shows the epoch of superluminal ejection, and the horizontal
solid line shows the standard deviation of the delay with respect to the
$\gamma$-ray flare; the dashed line connects the points of successive minima
and the maximum of the polarized radio flux density.}
\end{figure}
\begin{figure}
\figurenum{3b}
\plotone{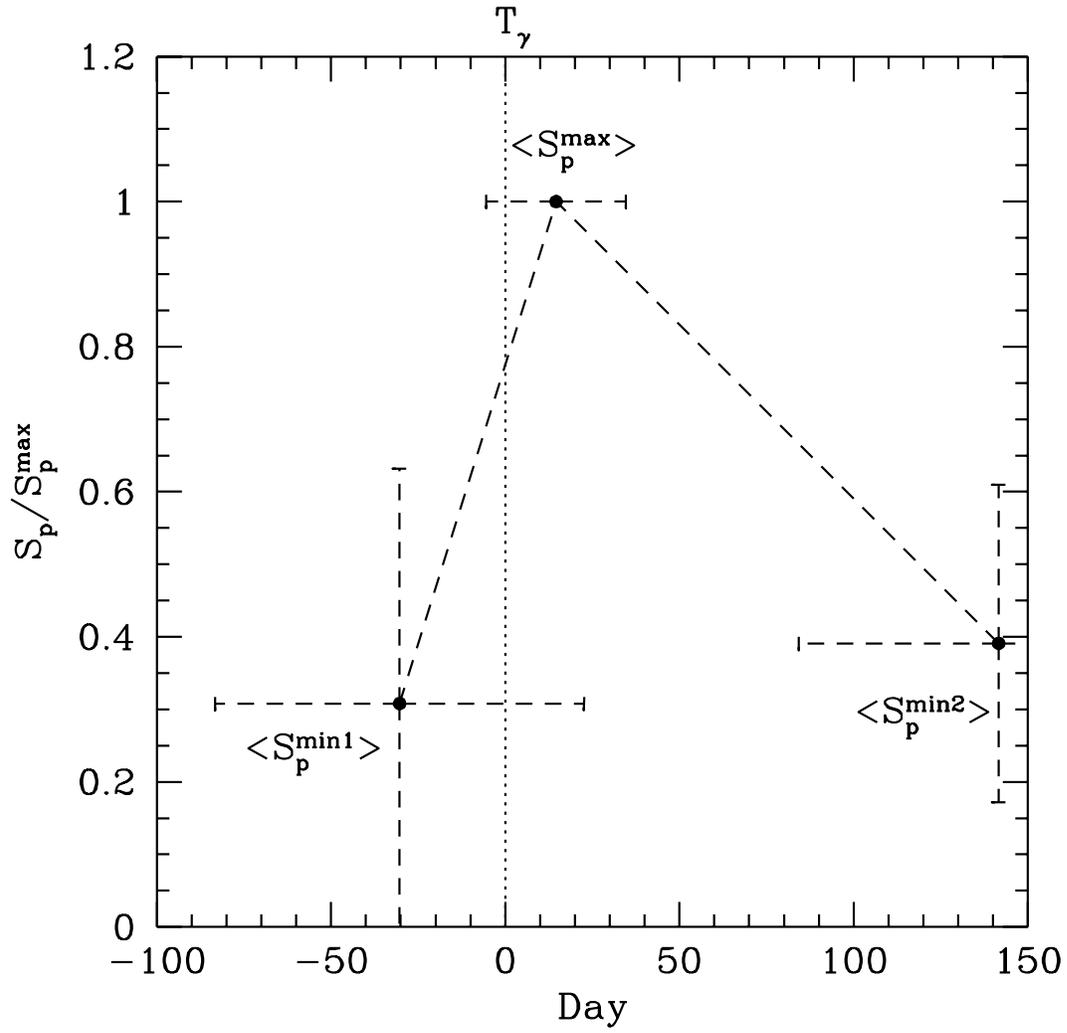}
\caption{Average temporal relationship between $\gamma$-ray flares,
and polarized radio flux densities; designations are as in Fig. 3a.}
\end{figure}
\end{document}